\documentclass[reprint, prl]{revtex4-2}
\usepackage{graphicx}
\usepackage{dcolumn}
\usepackage{bm}
\usepackage{hyperref}

\begin{document}

\title{Dual Operation of Gate-All-Around Silicon Nanowires at Cryogenic Temperatures: FET and Quantum Dot}

\author{C. Rohrbacher$^1$, J. Rivard$^1$, R. Ritzenthaler$^2$, B. Bureau$^1$, C. Lupien$^1$, H. Mertens$^2$, N. Horiguchi$^2$ and E. Dupont-Ferrier$^1$}
\affiliation{
$^1$Institut quantique and Département de Physique,
Université de Sherbrooke, Sherbrooke, Québec, J1K 2R1, Canada\\
$^2$IMEC, Kapeldreef 75, 3001 Leuven, Belgium.\\
}

\date{\today}

\begin{abstract}
As CMOS structures are envisioned to host silicon spin qubits, and for co-integrating quantum systems with their classical control blocks, the cryogenic behaviour of such structures need to be investigated. In this paper we characterize the electrical properties of Gate-All-Around (GAA) n-MOSFETs Si nanowires (NWs) from room temperature down to 1.7 K. We demonstrate that those devices can operate both as transistor and host quantum dots at cryogenic temperature. In the classical regime of the transistor we show improved performances of the devices and in the quantum regime we show systematic quantum dots formation in GAA devices.
\end{abstract}

\maketitle
\section{Introduction}

Silicon based technology has shown great potential for quantum information processing\cite{Zwanenburg2013}. Silicon spin qubits have proven to be competitive solid-state qubit system with very long coherence time\cite{Muhonen2014,yoneda2018quantum}, and single and two-qubit gate fidelity above the error correction threshold has been demonstrated\cite{Yang2019,lawrie2021simultaneous}. The compatibility with CMOS technology allows leveraging the capabilities of the microelectronics industry to built large-scale quantum systems within industrial foundry lines. 

With quantum systems scaling up, co-integrating electronics at low temperature for qubit readout and control has become a necessity\cite{VanDijk2019,Charbon2017}. Silicon systems allow for building the control electronics and quantum processors on the same chip and a few architectures have been proposed in that direction\cite{Vandersypen2016,Veldhorst2017} motivating the development of CMOS devices operating from 4 K to sub-kelvin temperature. Studying cryogenic performance of CMOS technology has therefore become highly relevant and efforts have recently been made to implement compact modelling down to few Kelvin  \cite{Beckers2018,Pahwa2021}. Hence, finding innovative, industry compatible, CMOS technology that grant possibilities for large scale integrated control electronics and can be used for quantum information processing is a significant milestone to reach. For now, spin qubit made with industry standard fabrication processes have only been demonstrated in Si-MOS with polysilicon gate\cite{Stuyck2021}, FD-SOI\cite{Maurand2016}, and FinFET \cite{Zwerver2021,Camenzind2022} technologies. 

Simultaneously, over the last decades, the semiconductor industry introduced a tremendous number of innovations (for the most notable strain engineering \cite{Bai2004,Mistry2007}, High-k/Metal gates (HKMG) introduction \cite{Mistry2007,Auth2008}, and  Gate-Last integration \cite{Mistry2007,Auth2008}) in order to pursue the device dimensions downscaling (necessary to improve the transistors’ performance). 
Additionally, the ‘Planar bulk transistor’ architecture itself faced serious physical problems at short gate lengths due to high channel doping and Short Channel control. Therefore, Multiple-gate transistors such as FinFETs emerged as solutions for the 22~nm node and below to provide pace for further downscaling due to their fully depleted nature ensuring an excellent short channel effects control \cite{Auth2012}.
Due to their optimal electrostatic control of the channel, Silicon nanowires/nanosheets  MOSFETs are the candidate device to extend the gate length and gate pitch scaling of MOSFET transistors beyond the FinFET limits \cite{Mertens2016,Mertens2017}. A lateral process offers the advantage of a process flow relatively comparable to FinFETs, and vertical stacking allows maximizing the drive current for a given footprint on the wafer.

In this letter we study the performance of GAA nMOSFETs silicon nanowires at cryogenic temperature and show functional device down to 60 mK with enhanced performance. 
Additionally, we observe systematic formation of quantum dots in those devices and investigate on their possible origins. We demonstrate electrostatically defined quantum dot formed in the nanowire channel which indicates that this technology is suitable for both 4K electronics and as quantum dot structures making GAA Nanowires (NWs) devices relevant for quantum information processing. 
                            
\section{Device fabrication and measurement setup}
\begin{figure}[htp]
  \includegraphics{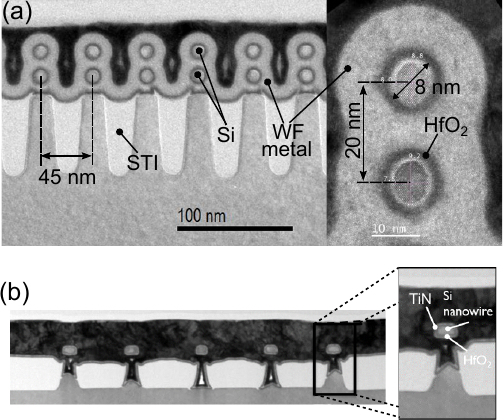}
  \caption{\label{fig:Wire} TEM transversal cross-section of nanowires used in this work, showing device dimensions and gate stack. 
  (a) The nanowires diameter is 8 nm, and the gate stack layers (Interfacial layer, high-k, Metal Gate and fill metal) are visible on TEM.
  (b) Single nanowires structures for quantum dot investigation from a relaxed dimensions maskset similar to the device measured. The nanowire measured has a width \mbox{W = 10 nm} and a gate length $\mathrm{L_G}$ = 20 nm. The Metal Gate in this case is purely TiN.}
  \end{figure}

This paper details cryogenic behaviour in the classical and quantum regime of a n-channel GAA NW MOSFET fabricated at IMEC. The process flow of the nanowires used in this work is shown in more detail in \cite{Mertens2016,Mertens2017}. Starting from a bulk silicon wafer, SiGe/Si epitaxial layers are grown and patterned. Dummy gate oxide and dummy gate are then deposited and patterned. Next, spacers are formed, and S/D epitaxial modules grown. Interlayer Dielectrics (ILD0) is then deposited, and a CMP (Chemical Mechanical Polishing) is performed in order to access the dummy gate. After the dummy gate removal, the nanowires are released using a selective SiGe etch. In order to form the final gate stack, gate dielectric and CMOS dual workfunction metal layers ($\mathrm{SiO_2}$ interfacial layer, $\mathrm{HfO_2}$ high-k, Work Function Metal Gate layers, and fill metal) are then deposited, resulting in a gate stack EOT (Equivalent Oxide Thickness) in the 1~nm range. The end of processing consists in standard back-end and metal deposition up to the Metal 1 level.
The transistors are made of four to twenty wires in parallel (Fig.\ref{fig:Wire}a-b), with a gate length $\mathrm{L_G}$ ranging from 24 nm up to 70 nm and a width W =  8 nm. A device with single wire with a width W = 10 nm and gate length $\mathrm{L_G}$ = 20 nm was also studied to investigate coulomb blockade in a single wire (Fig.\ref{fig:Wire}c).

I-V characteristics at temperatures from 300 K to 1.7 K were performed in a Janis variable temperature insert cryostat. Single wire transistors were measured in a bottom loading Bluefors dilution fridge operating at 12 mK base temperature and a 60 mK sample temperature. Conductance measurement was made by standard lock-in detection.

\section{Transistor behaviour at low temperature}
\begin{figure}[htp]
  \includegraphics{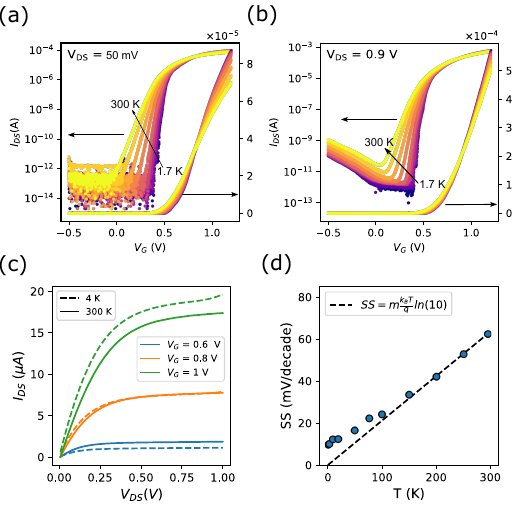}
  \caption{\label{fig:I-V} I-V characteristics taken for a $L_G$ = 70 nm device with 20 NWs in parallel taken in linear (a) and saturation regime (b). (c) Output characteristics at 300 K and 4 K taken at fixed gate voltage. (c) Evolution with temperature of SS. In dotted black is plotted the standard model $SS = m\frac{k_BT}{q}\mathrm{ln10}$}
  \end{figure}

Typical transfer and output characteristics of a device with 20 NWs in parallel are shown in Fig.\ref{fig:I-V} showing functional transistors down to 1.7 K. At low temperature, we notice a slight increase in $I_{ON}$ due to higher mobility because of the decrease of phonon scattering. At the same time, we observe a significant 76\% decrease of $I_ {OFF}$ from 300 K to 1.7 K indicating improved leakage performance at cryogenic temperatures.  Threshold voltage is extracted using linear extrapolation method\cite{Ortiz-Conde2002}. We show an increase of 100 mV from 300 K to 1.7 K caused by the shift of the Fermi level with temperature\cite{Beckers2020VTH}. 

We extracted subthreshold swing (SS) in linear regime with respect to temperature as shown in Fig.\ref{fig:I-V}.d. The SS drops from 62.2 mV/decade at 300 K down to 10.1 mV/decade at 1.7 K demonstrating an improvement in switching performance. Basic SS(T) model is defined by: $SS = m\:\mathrm{ln10}\:k_BT/e$, where $ m = (C_{ox} + C_{it} )/C_{ox} $, $C_{it}$ being the interface trap capacitance and $C_{ox}$ the oxide capacitance. This model shows linear temperature dependence at high temperature and a saturation of SS below 20 K (see Fig\ref{fig:I-V}.d). This phenomenon has been universally observed in a wide range of CMOS devices \cite{Incandela2017,Bohuslavskyj2016,Cretu2016,Han2021, Galy2018} including in Si GAA NWs\cite{Boudier2018}. It has been recently explained by considering a disordered-induced band tail at the conduction band edge \cite{Beckers2020} though the nature of disorders remains vague. This consideration has been successfully modelled within the BSIM framework \cite{Pahwa2021} by using an effective temperature $T_0$ defined as a temperature cutoff: $SS_{T\ll T_0}= m\:\mathrm{ln10}\:k_BT_0/e$ \cite{Bohuslavskyi2019}. Here, taking m = 1.07 from the SS at 300 K we extract $T_0 = \mathrm{48} \pm \mathrm{2.8\:K}$. 

\section{Quantum dot observation}

\begin{figure}[htp]
  \includegraphics{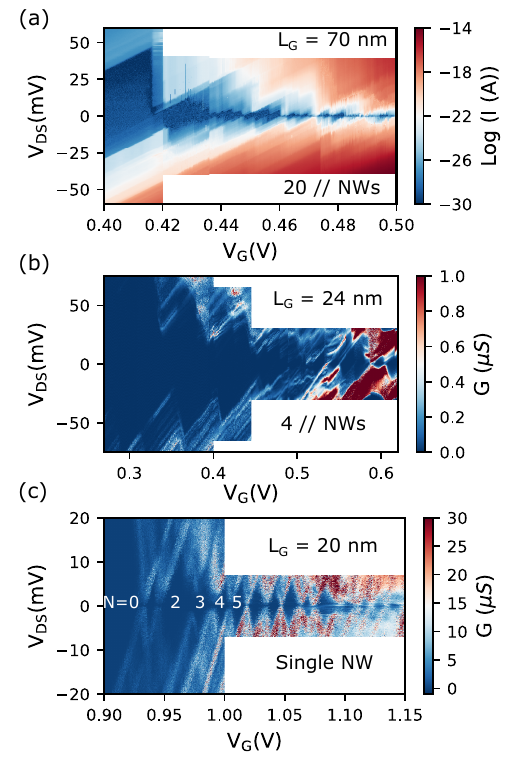}
  \caption{\label{fig:dot} Stability diagrams from three different devices. (a-b) Stability diagram of devices with gate length $L_G$ = 70 nm and $L_G$ = 24 nm made with respectively 20 and 4 NWs taken at T = 1.7 K. (c) Stability diagram of a $L_G$ = 20 nm device with a single wire taken at 60 mK. Electron filling number are indicated.}
  \end{figure}

\begin{figure}[htp]
  \includegraphics{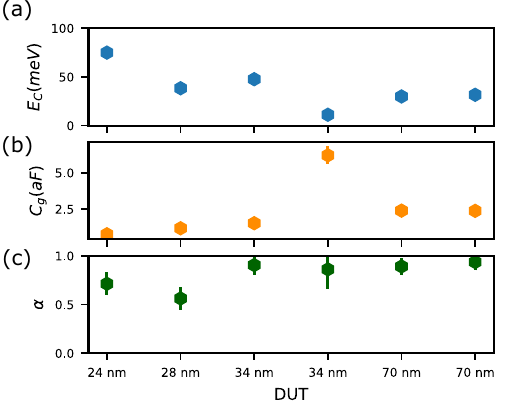}
  \caption{\label{fig:param} Distribution of the charging energy $E_C$ (a), the gate capacitance $C_G$ (b) and the lever arm $\alpha = \frac{C_G}{C_{\Sigma}}$ (c) for all  measured NW's identified by gate length. Data are all extracted only from the first coulomb diamond. Device with $L_G$ = 24, 28 and 34 nm are made of 4 NWs, device with $L_G$ = 70 nm are made with 20 NWs.}
  \end{figure}

 We then investigate the behaviour of GAA devices below threshold and performed a systematic study for wires with several gate lengths : 24, 28, 34 and 70 nm. Coulomb blockade oscillations were observed for low $V_{DS}$ on all devices, indicating the presence of quantum dots. Two examples of typical stability diagrams measured for devices of gate length 70 nm and 24 nm are shown in Fig.\ref{fig:dot}a-b. The several superpositions of diamonds observed is a result of multiple parallel transport paths which is expected due to the 4 (Fig\ref{fig:dot}b) to 20 (Fig\ref{fig:dot}a) parallel NWs   \cite{Nilsson2016}.We restrict therefore our analysis to the first few diamonds where transport is dominated by a reduced number of wires. From the Coulomb diamonds shape, we can extract relevant parameters : the charging energy $E_C = \frac{e^2}{C_{\Sigma}}$, corresponding to the energy required to add an electron to the dot with $C_{\Sigma}$ being the total capacitance to the quantum dot. The gate capacitance to the dot $C_G$ and the lever arm defined as $\alpha = \frac{C_G}{C_{\Sigma}}$. A systematic analysis and comparison of key parameters : Charging energy (Ec), lever arm (alpha)  and gate capacitance (Cg) for devices of various gate length 24 nm, 28,34 and 70nm is shown in Fig\ref{fig:param}. Charging energy among devices exhibits a distribution ranging from 75 meV to 11 meV indicating a very strong confinement. The high value of the charging energy, the fact that it does not scale with NWs dimensions and the impossibility to see regularly spaced Coulomb oscillations indicate that these dots are not electrostatically defined in the channel but rather unintentional dot. This parasitic dot can arise from various causes and have been a common observation in several types of quantum dot devices \cite{Thorbeck2012,Nordberg2009,Mason2004}. Causes for unintentional dots include strain induced either by the oxide or the metal gate, that causes modulation the bandgap strong enough to trap charges \cite{Thorbeck2012,Thorbeck2015}. Interface trap and defect charges causing random potential modulation can generate parasitic dots and increase electron scattering \cite{Nordberg2009}. The excellent lever arm $\alpha$ measured in the range from 0.56 to 0.93 suggests that the dots are strongly coupled to the gate thanks to the GAA geometry. Nevertheless, the wide distribution of the lever arm might indicate that the unintentional dots have various origins. Unfortunately, several NWs in parallel hinder the analysis and increase the probability of probing different dots caused by multiple disorder, hence reducing the chance of measuring electrostatic quantum dot as well as inferring on the nature of the unintentional dots. These devices can nonetheless be used for classical electronics for their excellent properties such as improved SS (see fig\ref{fig:I-V}). For quantum application one would need to use the same technology with a single NW. 

To further explore the NW's behaviour in the quantum regime, a device with \textit{single} NW of dimension $\mathrm{L_G}$ = 20 nm and D = 10 nm was measured at low temperature, down to 60 mK for improved resolution. The stability diagram visible in Fig.\ref{fig:dot}.c shows an improvement in diamonds visibility as expected from diminishing the number of transport paths. We observe a regular pattern of Coulomb diamonds from the single electron regime up to $N_{electron} \approx $ 13  where the transistor starts to open. From the regular-sized Coulomb diamonds we extract $E_C$ = 3.5 $\pm$ 0.6 meV, gate capacitance, $C_G$ = 11.8 $\pm $1.2 aF and lever arm $\alpha$= 0.26 $\pm $ 0.06. The lever arm factor is smaller than the previous design as the quantum dot is formed in the channel, yet the value of the lever arm factor is on par with very good electrical control in the channel. Using a very simple cylindrical capacitive model and fixing L = 20 nm, we derive a dot size of 6 nm diameter which is in good agreement with the device dimensions, a strong indication that the dot is caused by controllable gate \mbox{modulation}. 

\section{Conclusion}

In conclusion, we have shown that silicon GAA NW are behaving as classical MOSFET down to millikelvin temperature with improved performances. Single nanowire study shows that the GAA architecture is hosting controllable quantum dot down to the single electron regime, a building block for spin qubit devices. This study shows that GAA devices are promising candidate for co-integrating quantum dot devices in silicon with control electronics using the same technological node. 

\section{Acknowledgement}

The authors would like to thank all the colleagues from the Institut quantique and Université de Sherbrooke for their technical support during the experimental process, and especially M. Lacerte. E.D-F acknowledge support from NSERC, the Canada First Research Excellence Fund and FRQNT.

\bibliography{QuantumTransportNWBib}

\end{document}